\documentclass[lettersize,journal]{IEEEtran}

\usepackage{placeins}
\usepackage{booktabs} 
\usepackage{tabularx}  
\usepackage{ragged2e}  
\usepackage{siunitx}    
\usepackage[T1]{fontenc}
\usepackage[utf8]{inputenc}

\usepackage{mathtools} 

\usepackage{amsmath, amssymb, amsfonts, amsthm, bm}

\usepackage{siunitx}  
\usepackage{textcomp} 

\usepackage{graphicx}
  \graphicspath{{./Figs/}}
\usepackage[caption=false,font=footnotesize]{subfig}

\usepackage{booktabs}
\usepackage{tabularx}
\usepackage{multirow}
\usepackage{array}       
\usepackage{rotating}    
\newcolumntype{P}[1]{>{\centering\arraybackslash}p{#1}}
\newcolumntype{M}[1]{>{\centering\arraybackslash}m{#1}}

\usepackage{enumitem}
  \setlist{noitemsep,topsep=3pt,parsep=2pt,partopsep=2pt}

\usepackage[dvipsnames]{xcolor}  
\usepackage{cite}
\usepackage{hyperref}

\colorlet{linkcol}{blue!60!black}  
\colorlet{citecol}{green!60!black} 
\colorlet{urlcol}{blue!60!black}   

\hypersetup{
  colorlinks=true,
  linkcolor=linkcol,
  citecolor=citecol,
  urlcolor=urlcol
}

\usepackage{algorithm}
\usepackage{algpseudocode}

\usepackage{xcolor}
\usepackage{tikz}
\usetikzlibrary{
  positioning,
  fit,
  backgrounds,
  shadows,
  arrows.meta,
  calc 
}

\definecolor{myred_base}{HTML}{440154}
\definecolor{myblue_base}{HTML}{31688E}
\definecolor{mygreen_base}{HTML}{21908C}
\definecolor{myyellow_base}{HTML}{FDE725}
\definecolor{mygray_base}{HTML}{555555}
\definecolor{mypanelbg}{HTML}{F5F5F5}

\colorlet{myblue}    {myblue_base!40!white}   
\colorlet{mygreen}   {mygreen_base!40!white}
\colorlet{myyellow}  {myyellow_base!40!white}
\colorlet{myred}     {myred_base!20!white}
\colorlet{mygray}    {mygray_base!90!white}

\colorlet{mypanelbg} {myyellow_base!5!white} 

\tikzset{
  baseblock/.style={
    rectangle,
    rounded corners=5pt,
    draw=mygray, 
    thick,
    align=center,
    font=\sffamily\small, 
    text=black,
    drop shadow={opacity=0.15, shadow xshift=1pt, shadow yshift=-1.5pt, fill=mygray} 
  },
  mainblock/.style={
    baseblock,
    minimum width=1.3cm,
    minimum height=1.2cm
  },
  endblock/.style={
    baseblock,
    minimum width=1.3cm,
    minimum height=1.2cm
  },
  flow/.style       = {-{Stealth[length=2.5mm, width=2mm]}, thick, draw=mygray},
  dashedflow/.style = {flow, dashed},
  panel/.style      = {
    rectangle, draw=mygray, fill=mypanelbg, rounded corners=8pt
  },
  panellabel/.style = {
    font=\sffamily\bfseries, text=mygray
  }
}
\usepackage{pifont}
\usepackage{threeparttable}

\newcommand{\vect}[1]{\bm{#1}}
\newcommand{\mat}[1]{\mathbf{#1}}
\newcommand{\abs}[1]{\lvert#1\rvert}

\newcommand{\cmark}{\ding{51}}
\newcommand{\xmark}{\ding{55}}

\usepackage{stfloats}     
\fnbelowfloat             

\begin{document}

\title{Statistically Adaptive Differential Protection for AC Microgrids Based on Kullback--Leibler Divergence}

\author{Shahab Moradi Torkashvand, Arina Kharazi, Emad Sadeghi,~\IEEEmembership{Member,~IEEE,}\\
Seyed Hossein Hesamedin Sadeghi,~\IEEEmembership{Senior Member,~IEEE,}
and Adel Nasiri,~\IEEEmembership{Fellow,~IEEE}%
\thanks{Open-source code, scripts, and full results:
\href{https://github.com/emadsadeghi/kl-differential-protection}{\texttt{Github Repository}}.}%
\thanks{S.~M. Torkashvand, A. Kharazi, and S.~H.~H. Sadeghi are with the Department of Electrical Engineering,
Amirkabir University of Technology (Tehran Polytechnic), Tehran 1591634311, Iran
(e-mails: \href{mailto:sh.torkashvand@aut.ac.ir}{sh.torkashvand@aut.ac.ir};
\href{mailto:arinakharazi@aut.ac.ir}{arinakharazi@aut.ac.ir};
\href{mailto:sadeghi@aut.ac.ir}{sadeghi@aut.ac.ir}).}%
\thanks{E. Sadeghi and A. Nasiri are with the Department of Electrical Engineering and Computer Science,
University of South Carolina, Columbia, SC 29208, USA
(e-mails: \href{mailto:ssadeghi@email.sc.edu}{ssadeghi@email.sc.edu};
\href{mailto:nasiri@email.sc.edu}{nasiri@email.sc.edu}).}%
}

\markboth{IEEE TRANSACTIONS ON SMART GRID,~Vol.~XX, No.~X, MONTH~YEAR}%
{Torkashvand \MakeLowercase{\textit{et al.}}: Statistically Adaptive Differential Protection for AC Microgrids}

\maketitle

\begin{abstract}
The proliferation of inverter-based resources challenges traditional microgrid protection by introducing variable fault currents and complex transient behaviors. This paper presents a statistically adaptive differential protection scheme based on Kullback–Leibler divergence, which is practically implemented via a Bartlett-corrected G-statistic computed on logarithmic transformed current magnitudes. The core of the method is a multivariate fault detection engine that employs the Mahalanobis distance to distinguish healthy and faulty states, enabling robust detection even in noisy environments, with detection thresholds statistically derived from a chi-squared distribution for precise control over the false alarm rate. Upon detection, a lightweight classifier identifies the fault type by assessing per-phase G-statistics against dedicated thresholds, enhanced by a temporal persistence filter for security. Extensive simulations on a modified CIGRE 14-bus microgrid demonstrate the scheme's high efficacy, achieving sub-cycle average detection delays and high detection and classification accuracy across various operating modes, high-impedance faults up to \SI{250}{\ohm}, \SI{10}{\milli\second} communication delay, and noise levels down to a \SI{20}{dB} signal-to-noise ratio. These findings highlight a transparent, reproducible, and computationally efficient solution for next-generation AC microgrid protection.
\end{abstract}

\begin{IEEEkeywords}
Microgrids, High-Impedance Faults, Differential Protection, Kullback–Leibler Divergence, Mahalanobis Distance.
\end{IEEEkeywords}

\section{Introduction}
\label{sec:introduction}
\IEEEPARstart{T}{he} global shift toward decentralized power generation, driven by environmental goals and energy security, has accelerated the growth of microgrids \cite{ref1}. These localized systems integrate distributed energy resources (DER), such as solar photovoltaics (PV), wind turbines (WT), battery energy storage systems (BESS), synchronous generators (SG), and combined heat and power (CHP) units with local loads, offering benefits like improved resilience, lower transmission losses, and greater consumer participation \cite{ref2}.

Microgrids can operate either grid-connected or in islanded mode \cite{ref3}. While this flexibility is advantageous, it poses significant challenges for conventional protection schemes \cite{ref4}. Reliable microgrid operation depends on effective protection strategies that can promptly manage faults and disturbances to maintain power supply for critical infrastructures \cite{ref5}.

The protection of microgrids is inherently challenging due to several distinguishing characteristics. First, fault current levels can vary significantly in grid-connected operation, the utility contribution may be substantial, while in islanded mode the short-circuit capacity is much lower and mainly determined by the characteristics of various types of DERs \cite{ref6}. Second, the growing integration of inverter-based resources (IBRs) leads to non-conventional fault responses. Unlike synchronous generators that sustain high fault currents, IBRs typically restrict their fault output to 1.1–2.0 per unit, often accompanied by complex transients, harmonics, and rapid phase shifts that conventional protection schemes may fail to address  ~\cite{ref6, ref7, ref8}. Third, frequent topology changes and bidirectional power flows in microgrids complicate the coordination of directional and overcurrent relays, thereby increasing the risk of maloperation or loss of selectivity  \cite{ref9}.

Conventional protection methods, such as overcurrent, distance, voltage-based, and impedance-based schemes, often prove inadequate in microgrid settings \cite{ref10}. The limited fault current levels, high penetration of inverter-based resources, and frequent topological changes challenge the assumptions of these methods and lead to issues such as miscoordination and decreased reliability  \cite{ref5}. While adaptive protection schemes have been explored to address these issues  \cite{ref11,ref12,ref13}, they frequently rely on complex real-time system analysis, extensive communication networks, and sophisticated coordination algorithms, which can introduce their own set of challenges related to cost, reliability, and cybersecurity.

Differential protection, by contrast, is inherently selective and resilient to external faults, load changes, and bidirectional flows, making it suitable for feeders, transformers, and DER interconnections \cite{ref14,ref15}. Yet its effectiveness is limited by the low fault currents of IBRs and the need for complex relay settings under diverse microgrid conditions \cite{ref16}. To overcome these challenges, advanced signal processing and statistical techniques have been proposed. Methods such as sign-based schemes \cite{ref17}, variational mode decomposition--Hilbert transform \cite{ref18}, Hilbert--Huang transform  \cite{ref19}, and empirical mode decomposition \cite{ref20} 
enhance directionality and nonlinear signal analysis but remain sensitive to noise and topology-specific tuning. Wavelet- and S-transform-based techniques \cite{ref21,ref22,ref23} provide strong time--frequency localization yet rely on empirical parameter selection, while Teager--Kaiser Energy Operator based methods \cite{ref24} offer robustness against load variations but degrade in high-impedance fault (HIF) scenarios.

Finally, statistical distance measures, in particular Kullback--Leibler (KL) divergence~\cite{ref25}, offer a comprehensive framework for comparing current signal distributions. KL divergence (also known as relative entropy) has demonstrated a strong capability in detecting subtle and HIF scenarios where traditional protection schemes often fail because of its high sensitivity to small distributional changes~\cite{ref26, ref28, ref29}. Moreover, KL divergence-based methods are known to perform relatively well even under moderate measurement noise, maintaining their effectiveness in complex, inverter-dominated microgrids. However, these schemes are not without limitations. The effectiveness of KL divergence-based protection can still be challenged under severe noise conditions or in the presence of extremely~HIF, where the distinction between healthy and faulty signal distributions may become less pronounced. More importantly, accurate and robust threshold selection remains critical; many existing implementations rely on fixed or empirically tuned thresholds that may not generalize well to varying operational scenarios, potentially reducing reliability and adaptability in real-world microgrid environments.

This work presents a differential protection scheme that leverages KL-divergence-inspired statistical change detection. Unlike heuristic thresholding, the proposed method employs the G-statistic (likelihood-ratio chi-square), which follows a well-defined chi-square ($\chi^2$) distribution under healthy conditions, thereby enabling controlled false-alarm rates. The approach uses the Mahalanobis distance on the three per-phase G-statistics to detect deviations while accounting for inter-phase correlations. Key contributions include: (a) Bayesian optimization to select adaptive histogram binning; (b) principled threshold derivation from the $\chi^2$ distribution; and (c) hierarchical classification with temporal persistence voting for noise immunity. This establishes a transparent, statistically sound framework for microgrid protection.

The manuscript is organized as follows. Section~II formulates the proposed differential protection method and describes threshold selection and fault classification. Section~III presents simulation results, demonstrating the method's performance in an AC microgrid under various case studies, including topological and mode changes as well as high-impedance faults. Finally, Section~IV compares the proposed method with state-of-the-art differential protection schemes, highlighting its advantages in detection time and robustness.

\section{Proposed Methodology}
\label{sec:proposed_methodology}
\begin{figure*}[t]
    \centering
    \includegraphics[width=\textwidth]{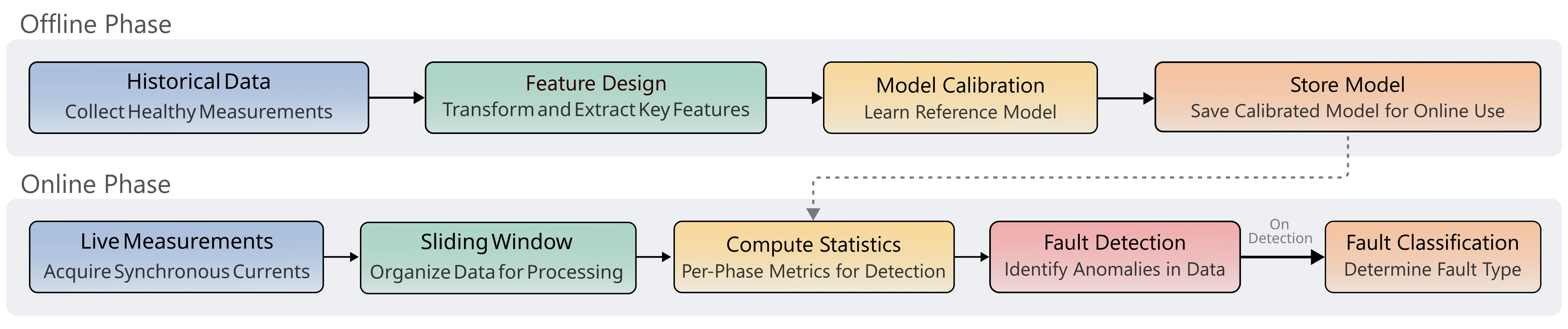}
    \caption{Workflow overview: \emph{Offline phase}—design features, calibrate the reference model, and store it for online use; \emph{Online phase}—compute per-phase statistics, perform fault detection, and classify faults.}

    \label{fig:method-flow}
\end{figure*}

A high-level workflow of the proposed method is shown in Fig.~\ref{fig:method-flow}. As can be seen in this figure, the methodology is structured into two distinct phases: (i)~an \textit {offline tuning phase} where a statistical model of the healthy system state is learned from historical data, and (ii)~an \textit{online operational phase} where real-time currents are continuously monitored for deviations from this healthy model.

\subsection{Offline Tuning Phase}
\label{subsec:offline-tuning}
The goal of the offline phase is to learn a robust statistical representation of the system’s healthy operating conditions. In constructing the healthy reference model, we considered all operating scenarios in which the protected line itself remains fault-free. This includes not only purely healthy network conditions but also events where faults occur on external lines. The rationale is that the protection scheme must remain robust to disturbances elsewhere in the system; therefore, the statistical baseline of the protected line should reflect its behavior under a wide range of system dynamics, provided the line under protection is healthy.

Referring to Fig.~\ref{fig:method-flow}, the offline tuning phase involves four main tasks. These include preparation of a representative simulated dataset, design of a statistical model to establish the features of the healthy system, calibration of the model, and storage of the essential parameters derived from the preceding stages for real-time deployment in the online operational phase.

\subsubsection{Historical Data}
The first step is to employ an appropriate simulation platform for constructing a datatset that characterizes the system’s normal behavior. This dataset serves as the baseline for identifying abnormal conditions. The raw input comprises three-phase current measurements at both the sending $(I_{a}^{s}, \; I_{b}^{s}, \; I_{c}^{s})$ and receiving $(I_{a}^{r}, \; I_{b}^{r}, \; I_{c}^{r})$ ends of the protected line. In addition to these six channels, the zero-sequence current component, defined as $\vect{I}_0 = \sum_p \vect{I}_p$, where $p \in \{a,b,c\}$, is also considered. Being primarily indicative of ground faults~\cite{ref24}, the $I_{0}$ signals from both the sending ($I_{0}^{s}$) and receiving ($I_{0}^{r}$) ends are appended as supplementary channels. This enriches the dataset and provides a more comprehensive statistical baseline for subsequent fault classification.

For each current set, $\bm{I}$, we have
\begin{equation}
\bm{I} = [I_{1}, I_{2}, \ldots, I_{n}, \ldots, I_{N}],
\label{eq:current_set}
\end{equation}
where $I_{n}$ denotes the $n$-th sample of the current waveform, and $N$ is the total number of samples in one cycle. This results in an eight-channel dataset.

To enhance the model’s robustness to measurement noise, small perturbations are added to the simulated current signals during training. This allows the model to learn features that are less sensitive to noise encountered during online operation, effectively acting as a form of data augmentation. Formally, if $\bm{I}\in\mathbb{R}^N$ denotes the current signal in a dataset, the augmented signal $\bm{I}^{\text{aug}}$ can be expressed as:

\begin{equation}
\label{eq:augmentation}
\bm{I}^{\text{aug}} \;=\; \bm{I} + \boldsymbol{\varepsilon},
\qquad
\boldsymbol{\varepsilon}\sim\mathcal{N}\!\big(\bm{0},\,\sigma^2\,\mat{I}_N\big)
\end{equation}

\noindent where $\boldsymbol{\varepsilon}$ is i.i.d.\ zero-mean AWGN and $\mat{I}_N$ is the $N{\times}N$ identity matrix. Note that the per-phase noise variance $\sigma^2$ is set to satisfy 
$\mathrm{SNR}_{\mathrm{dB}} = 10\log(I_{\mathrm{rms}}^2 / \sigma^2)$,
, where $\textit{I}_{\mathrm{rms}}^2$ represents the average power of the current signal under normal (fault-free) operating conditions.

After noise augmentation, the continuous current signals are partitioned into overlapping temporal windows to facilitate statistical analysis. Each window contains $L$ samples, and consecutive windows are separated by a step size~$S$. The complete collection of windows, ${W}$, is defined as:

\begin{equation}
    \label{eq:window-set}
    {W} = \{ \vect{w}_1, \vect{w}_2, \dots, \vect{w}_M \}
\end{equation}

\noindent where $M$ denotes the total number of windows extracted from the time series. Finally, each windowed current is further stabilized by a logarithmic transformation to mitigate variance and suppress the impact of outliers:
\begin{equation}
    \label{eq:log-transform}
    \tilde{\bm{I}} = \ln\left(1 + \abs{\bm{I}^{aug}}\right)
\end{equation}

The addition of 1 ensures numerical stability near zero, while the log operation compresses the dynamic range and symmetrizes the distribution. This step yields more consistent inputs for feature extraction. At this stage, the dataset is fully prepared for feature design.

\subsubsection{Feature Design}
\label{subsubsec:feature-design}
The protection scheme evaluates discrepancies between the statistical distributions of the sending and receiving currents. Instead of relying on simple amplitude comparisons, it employs the G-statistic (likelihood-ratio chi-square), to rigorously quantify these differences. Initially, for each window $\vect{w}_n$ and for each phase~$p$, histograms of the transformed signals, $\tilde{\bm{I}}_p$, are constructed. The respective G-statistic, $g_p$, is then computed as~\cite{ref29}:
\begin{equation}
    \label{eq:g-statistic}
    g_p = 2 \sum_{i\in{K}_p} \!\left[ n_{i}^{s}\,\ln\!\left(\frac{n_{i}^{s}}{e_i}\right)
                                            + n_{i}^{r}\,\ln\!\left(\frac{n_{i}^{r}}{e_i}\right) \right]
\end{equation}
where $K_p$ is the set of populated bins, $n_{i}^{s}$ and $n_{i}^{r}$ denote the observed counts in the $i$-th bin for the sending and receiving signals, respectively, and $e_i$ represents the expected count under the null hypothesis (healthy network condition) is:
\begin{equation}
    \label{eq:expected-count}
    e_i = \frac{n_{i}^{s} + n_{i}^{r}}{2}
\end{equation}

While $g_p$ provides a rigorous measure of distributional divergence, its finite-sample
behavior can deviate from the theoretical $\chi^2$ law, especially for short windows.
To mitigate this small-sample bias, we adopt the Bartlett correction~\cite{ref30}, yielding
the adjusted statistic $g_p^\star$:
\begin{subequations}\label{eq:bartlett}
\begin{align}
g_p^\star &= C_{B,p}\, g_p, \label{eq:bartlett_gstar}\\
\shortintertext{where}
C_{B,p} &= \left( 1 + \frac{K_p^{\mathrm{eff}} + 1}{6(2L - 1)} \right)^{-1} .
\label{eq:bartlett_cb}
\end{align}
\end{subequations}

\noindent
Here, $K_p^{\!\mathrm{eff}}$ denotes the number of populated histogram bins for phase $p$, bins for which at least one terminal (sending or receiving) has a strictly positive count. Only such non-empty cells are included in \eqref{eq:g-statistic}. This adjustment ensures that, under healthy network conditions, $g_p^\star$ more closely
follows its theoretical $\chi^2_{\,K_p^{\!\mathrm{eff}}-1}$ distribution \cite{ref30}, thereby stabilizing
the false-alarm rate.

Finally, the adjusted statistics from all three phases under healthy conditions are collected to form the G-statistic feature vector for each time window:
\begin{equation}
    \label{eq:g-feature-vector}
    \bm{g}_{h}^* = \left[ g_a^*, g_b^*, g_c^* \right]^\top
\end{equation}

This three-dimensional feature vector encapsulates the statistical discrepancy across all phases under healthy operation and serves as the primary input for the subsequent fault detection and classification stage.

\subsubsection{Model Calibration}
\label{subsubsec:model-calibration}
The representational power of the G-statistic strongly depends on histogram discretizations. Fixed-width binning is unsuitable for current data, as it obscures dense regions and amplifies tail noise. To address this, we use quantile-based adaptive binning, which partitions values according to healthy-signal quantiles. This yields finer resolution in high-density ranges and coarser bins in the tails, producing a more robust and informative feature space for statistical analysis.

Simultaneously, for all phases, the adaptive binning strategy is governed by three hyperparameters: the total number of bins ($K_p$) and two ratios ($r_{1,p}, r_{2,p}$) that allocate bins to the lower tail, central region, and upper tail \cite{ref31}. These hyperparameters, $\{K_p, r_{1p}, r_{2p}\}$, are tuned using Bayesian optimization \cite{ref32} to maximize a phase-wise Q--Q alignment objective on the healthy per-phase statistics:
\begin{gather}
\mathcal{O}=\tfrac{1}{3}\sum_{p\in\{a,b,c\}} \rho_p^{2},
\label{eq:objective-g}\\
\shortintertext{where}
\rho_p=\operatorname{corr}\!\big(
  \operatorname{sort}(g_p^\star),\,
  Q_{\chi^2}(K_p^{\mathrm{eff}}{-}1)
\big).
\label{eq:objective-rho}
\end{gather}

\noindent Here, $g_p^\star$ is the Bartlett-corrected G-statistic for phase $p$, $K_p^{\!\mathrm{eff}}$ is the number of populated histogram bins, and $Q_{\chi^2}$ denotes the $\chi^2$ quantile function \cite{ref33}.

\subsubsection{Essential Parameters Storage}
\label{subsubsec:Essential-Parameters-Storage}
The adjusted G-statistic vectors from the healthy training data,
$\bm{g}_{h}^{*}=\big[g_a^{*},\,g_b^{*},\,g_c^{*}\big]^\top$, define the
normal-operation distribution via its mean vector, $\bm{\mu}$, and covariance
matrix, $\bm{\Gamma}$:
\begin{subequations}
\label{eq:stats}
\begin{align}
\bm{\mu} &= \mathbb{E}\!\left[\bm{g}_{h}^{*}\right], \label{eq:mean_vector}\\
\bm{\Gamma} &= \mathbb{E}\!\left[
\big(\bm{g}_{h}^{*}-\bm{\mu}\big)\big(\bm{g}_{h}^{*}-\bm{\mu}\big)^\top
\right]. \label{eq:Gamma}
\end{align}
\end{subequations}

\noindent The diagonal elements of $\bm{\Gamma}$ are the per-phase variances of
$g_p^\star$, and the off-diagonal entries capture inter-phase linear
correlations. Exploiting these correlations in the detector enhances
sensitivity; treating phases independently would discard this information.

The offline tuning phase concludes with storage of the essential components of
the model: the mean vector $\bm{\mu}$, the covariance matrix $\bm{\Gamma}$, and the optimized adaptive histogram
edges $\{\xi_{k,p}\}$ for each phase $p$. These elements constitute a
data-driven statistical model of the protected line under healthy conditions
for deployment in real-time monitoring.

\subsection{Online Operation Phase}
\label{subsec:online-operation}

In the online phase, the protection scheme (i.e., real-time fault detection and classification)
operates in a continuous loop, processing incoming currents in sliding windows. Each window
performs three steps: phase-wise anomaly scoring, multivariate detection, and post-detection classification.

\subsubsection{Phase-wise anomaly scoring}
\label{subsubsec:online-scoring}
At each window $\vect{w}_n$, for each phase $p\in\{a,b,c\}$, we test whether the sending, $\pi^{s}_{p}$, and receiving, $\pi^{r}_{p}$, terminal
distributions, built with the fixed edges $\{\xi_{k,p}\}$ from the offline stage, are equal. This can be conceptualized by comparing the respective null hypothesis, $H_{0,p}$, representing the healthy condition, and the alternative hypothesis, $H_{1,p}$, defining the faulty condition, \textit{i.e.},
\begin{equation}
H_{0,p}:\ \pi^{s}_{p} = \pi^{r}_{p}
\quad \text{vs.} \quad
H_{1,p}:\ \pi^{s}_{p} \neq \pi^{r}_{p}
\label{eq:hypothesis}
\end{equation}

Here, the healthy condition corresponds to the case where the $\pi^{s}_{p}$ and $\pi^{r}_{p}$ are statistically identical, whereas the faulty condition indicates these two distributions differ significantly.

The distributional difference between $\pi^{s}_{p}$ and $\pi^{r}_{p}$ is quantified by the adjusted $G$-statistic in
\eqref{eq:g-statistic}-\eqref{eq:bartlett}, yielding a per-phase anomaly scoring index for
fault detection. The individual per-phase statistics are finally aggregated into a single
feature vector for the window ending at time $t$, defined as $\bm{g}^{*}(t) = [\,g_a^{*},\, g_b^{*},\, g_c^{*}] ^ \top\,$.

\subsubsection{Fault Detection}
\label{subsubsec:Fault-Detection}
Fault detection is performed by measuring the statistical deviation of $\bm{g}^*(t)$ from the
learned healthy model using the squared Mahalanobis distance, $D_{\!M}^2$~\cite{ref34}, \textit{i.e.},
\begin{equation}
\label{eq:mahalanobis-distance}
D_{\!M}^2(t) = \big(\bm{g}^*(t) - \bm{\mu}\big)^\top \bm{\Gamma}_{\lambda}^{-1} \big(\bm{g}^*(t) - \bm{\mu}\big)
\end{equation}

To ensure numerical stability in computing the inverse covariance matrix $\bm{\Gamma}_{\lambda}^{-1}$,
Tikhonov regularization \cite{ref34} is applied. 
Instead of directly inverting $\bm{ \Gamma}$, 
the inverse of the regularized matrix is computed as:
\begin{equation}
\bm{\Gamma}_{\lambda}^{-1}= (\bm\Gamma + \lambda \textbf{I} )^{-1},
\label{eq:tikhonov}
\end{equation}
where $\textbf{I}$ is the identity matrix and \( \lambda > 0 \) is a small regularization parameter.
This improves conditioning and stabilizes the $D_{\!M}^2$ calculation.
 
Under the null hypothesis, assuming the G-vectors are approximately multivariate normal, $D_{\!M}^2$ follows a $\chi^2$ distribution with $p$ degrees of freedom ($p=3$ in our case). Large $D_{\!M}^2$ values indicate the observation is an outlier relative to
the healthy distribution.

The threshold for fault detection is derived from the $\chi^2$  distribution to correspond to a specified false alarm probability $\alpha_{\text{det}}$, thereby regulating the sensitivity of the detection process. 
A fault is detected if $D_{\!M}^2$ exceeds a pre-defined threshold, $\tau_{\text{det}}$. For a 3-phase system ($p=3$), the threshold is calculated as the inverse cumulative distribution function (CDF)
of the \(\chi^2_3\) distribution \cite{ref34}:
\begin{equation}
\label{eq:detection-threshold}
\tau_{\text{det}} = F^{-1}_{\chi^2_3}\!\big(1 - \alpha_{\text{det}}\big)
\end{equation}
A trip signal is initiated at the end of the first window for which $D_{\!M}^2(t) >
\tau_{\text{det}}$. 

\subsubsection{Fault Characterization}
\label{subsubsec:Fault Characterization}
Once a fault is detected, a classification stage is invoked to identify the specific type of fault. This stage reuses the computed per-phase statistics, $g_p^*(t)$, to determine which phases are involved. For each phase $p$, a standardized $z$-score, $z_p(t)$, is calculated~\cite{ref34}:
\begin{equation}
    \label{eq:z-score}
    z_p(t) = \frac{g_p^*(t) - \mu_p}{\sigma_p}.
\end{equation}
where \( \mu_p \) and \( \sigma_p \) are the per-phase mean and standard deviation
learned from the healthy training data. A phase $p$ is flagged as potentially faulted if its statistic, $g_p^*(t)$, satisfies either of the following two conditions.

\noindent\textit{(i) Absolute test:}
\begin{equation}
    \label{eq:absolute-test}
    \abs{z_p(t)} > z_p^{\mathrm{cls}},
\end{equation}
where \( z_p^{\mathrm{cls}} = Z^{-1}(1 - \alpha_p^{\mathrm{cls}}/2) \) is the phase classification threshold derived from the standard normal distribution for a given false alarm rate, $\alpha_p^{\mathrm{cls}}$. Notice that $Z^{-1}$ denotes the inverse standard-normal CDF/quantile.

\noindent\textit{(ii) Relative jump test:}
\begin{equation}
    \label{eq:relative-jump-test}
    \abs{g_p^*(t) - g_p^*(t-S)} > \Delta G_{p}^{\text{cls}}
\end{equation}
where, $\Delta G_{p}^{\text{cls}}$ is a pre-defined threshold for the change in the $G$-statistic between two consecutive windows (separated by hop $S$). This test is designed to capture abrupt increases indicative of a fault, even if the absolute $z$-score has not yet exceeded its threshold.

To enhance robustness against transient noise, these instantaneous per-phase flags are then filtered by a \textit{j}-of-\textit{m} temporal persistence vote \cite{ref34}. The final fault type is determined by the combination of phases whose flags persist after this vote. Although the computed $z$-scores vary across phases due to different $\mu_p$ and $\sigma_p$, the same absolute threshold is uniformly applied.

A similar two-test logic is applied to the zero-sequence statistic, $g_0^*(t)$, to detect ground involvement. Specifically, $g_0^*(t)$ is compared to its own thresholds: an absolute gate:
\begin{equation}
\tau_0^{\mathrm{cls}} = F^{-1}_{\chi^2_{\,K_{0}-1}}\!\big(1 - \alpha_0^{\mathrm{cls}}\big)
\label{eq:tau0_cls}
\end{equation}
derived from the $\chi^2$ distribution with $K_0{-}1$ degrees of freedom (where $K_0$ is the optimized bin count for the zero-sequence dataset), and a relative jump gate $\Delta G_{0}^{\text{cls}}$ (a predefined threshold on successive-window change). The resulting ground flag is also subjected to the same \textit{j}-of-\textit{m} persistence vote.

Finally, the vector of persistent phase/ground flags maps deterministically to a fault type: a single active phase+ground flag indicates a line-to-ground fault (\textit{ag, bg,} or \textit{cg}); two simultaneous phase flags indicate a line-to-line fault, either with ground (\textit{abg, acg,} or \textit{bcg}) or without (\textit{ab, ac,} or \textit{bc}); and three active phase flags indicate a three-phase fault (\textit{abc}).

\section{Results and Discussion}
\label{sec:results}
In this section, we present a comprehensive evaluation of the proposed fault detection and classification scheme. The analysis is structured to demonstrate the method's performance under a range of challenging yet realistic operating conditions, including network topology changes, HIFs in the presence of severe noise, dynamic load and generation shifts, and communication link delays. Through a combination of time-domain analysis of representative events and aggregated quantitative metrics, we validate the algorithm's sensitivity, accuracy, and robustness.

\begin{figure}[!t]   \centering
\includegraphics[width=1\columnwidth]{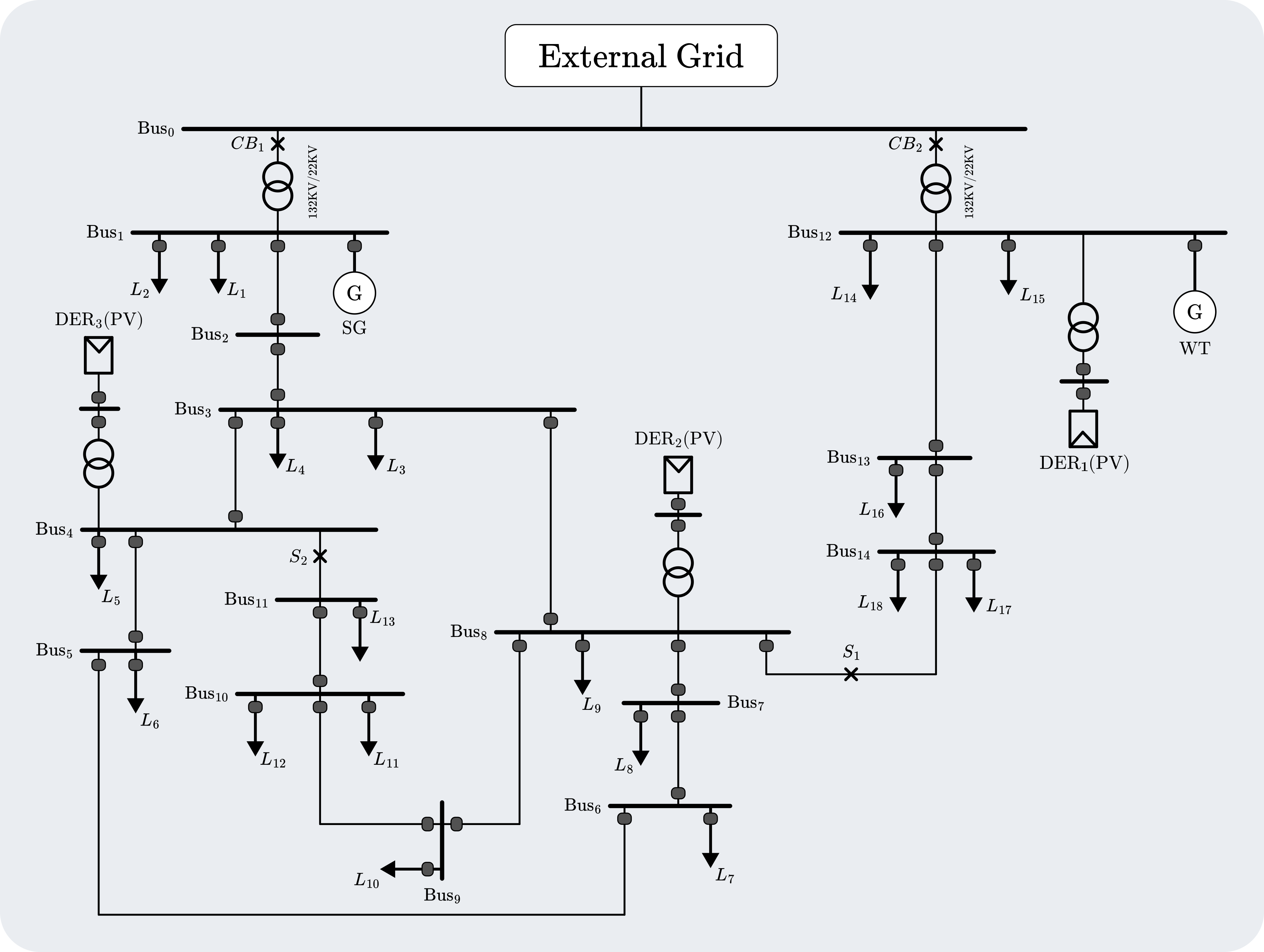}  
\caption{Single-line diagram of the modified CIGRE 14-bus microgrid test system.} 
\label{fig:single_line_diagram_final_v3}
\end{figure}

The proposed protection scheme is evaluated on a modified version of the CIGRE 14-bus benchmark low-voltage distribution network, adapted to represent a contemporary AC microgrid. This system, depicted in Fig.~\ref{fig:single_line_diagram_final_v3}, includes a mix of conventional loads and DERs for islanded operation capabilities. This microgrid operates at a voltage of \SI{22}{kV} and a frequency of \SI{50}{Hz}. Detailed data regarding the lines, loads, DER units, and transformers within the microgrid can be found in~\cite{ref35}.

The simulation results of all case studies were produced in DIgSILENT PowerFactory. Waveforms were simulated with a solver time step of \SI{10}{\micro\second} and sampled by the relay at \SI{10}{kHz}. The online detector operated with windows of length $L = 200$~samples (\SI{20}{ms}) and hop $S = 20$~samples (\SI{2}{ms}), i.e., 90\% overlap.

The scenarios span standard distribution fault categories, single-line-to-ground (LG), line-to-line (LL), double-line-to-ground (LLG), and three-phase (LLL), applied across a range of locations. Internal faults were placed at 20\%, 50\%, and 70\% of each protected line length. External faults on adjacent lines (sharing a bus) were applied at 1\% or 99\% from the common bus, while external faults on non-adjacent lines were located at 20\%, 50\%, and 70\% of the line. The resistance to faults ranged from \SI{0.1}{\ohm} to \SI{250}{\ohm}, ensuring that the model was trained under a wide range of fault conditions for improved generalization.

\subsection{Statistical Model Training and Thresholding Results}
\label{subsec:training-results}
Using the proposed method described in Section~\ref{sec:proposed_methodology}, a dedicated statistical model was trained for each protected line using only healthy operating data, with the offline tuning results summarized in Table~\ref{tab:offline_tuning_summary}.  As can be seen in this table, the optimized parameters, including the total number of bins ($K_p$) and the zone ratios ($[r_{1,p} : r_{2,p} : r_{3,p}]$) for three phases, and the number of bins ($K_0$) and the zone ratios ($[r_{1,0} : r_{2,0} : r_{3,0}]$) for zero-sequence current, vary across the different lines, reflecting the unique characteristics of each line's current distribution and leading to tailored decision spaces. Also, the per-phase mean ($\mu_p$) and standard deviation ($\sigma_p$), learned from healthy training data for each line are given in Table~\ref{tab:offline_tuning_summary}.

For a stringent $\alpha_{\text{det}} = 10^{-8}$, a uniform detection setting of $\tau_{\text{det}} = 40.13$ is applied for all lines. Similarly, with an $\alpha_p^{\text{cls}} = 10^{-8}$, a uniform phase classification thresholds, $z_0^{\text{cls}} = 5.73$, and $\Delta G_{p}^{\text{cls}} = 5$ are applied for all lines in the studied microgrid. While $\alpha_0^{\text{cls}} = 10^{-8}$, and $\Delta G_{0}^{\text{cls}}$ equals to $5$,  the $\tau_{0}^{\text{cls}}$ is not uniform and varies for each line, as can be seen in Table~\ref{tab:offline_tuning_summary}, where they are derived from the line-specific tuned models, and depend on $K_{0}$. Finally, for the \textit{j}-of-\textit{m} persistence vote, we used a 2-out-of-3 criterion (\textit{j}=2, \textit{m}=3). The corresponding calibration metric $\rho^{2}$, representing the average goodness of fit across all phases for each line, is also reported in Table~\ref{tab:offline_tuning_summary} for completeness.

\begin{table*}[t]
\centering
\renewcommand{\arraystretch}{1.1} 
\caption{Summary of Per-Line Offline Tuning Parameters.}
\label{tab:offline_tuning_summary}

\newcolumntype{S}{>{\centering\arraybackslash}p{1cm}}   
\newcolumntype{M}{>{\centering\arraybackslash}X}          
\newcolumntype{W}{>{\centering\arraybackslash}p{2.5cm}}    

\setlength{\tabcolsep}{3pt}

\begin{tabularx}{\textwidth}{@{} l M M W W S W S S @{}} 
\toprule
\textbf{Line ID} &
\bm{$K_{p}$} & \bm{$[r_{1,p}\!:\!r_{2,p}\!:\!r_{3,p}]$} & \bm{$[\mu_{a} : \mu_{b} : \mu_{c}]$} &
\bm{$[\sigma_{a} : \sigma_{b} :\sigma_{c}]$} &
\bm{$K_{0}$} & \bm{$[r_{1,0} :  \!r_{2,0}: \!r_{3,0}]$} & $\tau_0^{\mathrm{cls}}$ & $\rho^2 (\%)$\\
\midrule
Line 01-02 & 19 &  [0.21 : 0.35 : 0.44] &  [5.20 : 5.49 : 6.07] & [4.97 : 4.99 : 5.64] & 9 &  [0.10 : 0.10 : 0.80]  & 53.20   & 93.62\\ 
Line 02-03 & 16 &  [0.20 : 0.10 : 0.70] & [2.04 : 1.98 : 3.08] &  [2.22 : 2.21 :  3.10] & 15 &  [0.52 : 0.10 : 0.37] & 66.03  & 88.66 \\
Line 03-04 & 16 &  [0.47 : 0.10 : 0.43] & [4.62 : 4.63 : 5.31] & [3.63 : 3.94 : 4.23] & 14 &  [0.33 : 0.20 : 0.47]  & 64.00   & 91.76\\
Line 03-08 & 19 &  [0.10 : 0.12 : 0.78] & [1.88 : 2.32 : 4.60] &  [2.27 : 2.61 : 4.31] & 15 &   [0.53 : 0.10 : 0.37] & 50.8   & 94.66 \\
Line 04-05 & 16 &  [0.47 : 0.10 : 0.43] & [4.91 : 4.80 : 5.79] &  [3.88 : 3.81 : 4.83] & 11 &  [0.41 : 0.10 : 0.49]  & 57.67  & 96.28\\
Line 05-06 & 16 &  [0.47 : 0.10 : 0.43] & [5.82 : 5.98 : 7.87] & [4.22 : 4.20 : 5.68] & 18 &  [0.23 : 0.64 : 0.13]  & 71.94   & 93.35\\
Line 07-08 & 36 &  [0.41 : 0.17 : 0.42] & [9.27 : 10.8 : 13.7] & [4.25 : 5.28 : 7.04] & 8 &  [0.47 : 0.38 : 0.15]  & 50.01    & 89.21\\
Line 08-09 & 19 &  [0.53 : 0.37 : 0.10] & [9.41 : 8.88 : 9.64] & [3.62 : 4.14 : 4.37] & 14 &  [0.33 : 0.20 : 0.47] & 64.00    & 89.94 \\
Line 09-10 & 16 &  [0.21 : 0.10 : 0.69] & [2.54 : 2.94 : 5.19] & [2.58 : 2.92 : 5.06] & 18 &  [0.10 : 0.14 : 0.76] & 71.93    & 94.48 \\
Line 10-11 & 19 &  [0.10 : 0.10 : 0.80] &  [2.06 : 1.74 : 4.72] & [2.49 : 2.15 : 5.15] & 8 &  [0.21 : 0.35 : 0.44] & 50.81    & 90.64 \\
Line 12-13 & 16 &   [0.18 : 0.31 : 0.51] & [2.81 : 3.69 : 4.39] & [2.63 : 3.44 :  3.57] & 20 &  [0.52 : 0.25 : 0.23]  & 75.73 & 95.02\\
Line 13-14 & 20 &  [0.10 : 0.30 : 0.60] & [3.32 : 3.72 : 5.57] & [3.10 : 3.19 : 4.67] & 9 &  [0.53 : 0.37 : 0.10] & 53.16     & 93.72 \\
\bottomrule
\end{tabularx}
\end{table*}

\begin{table*}[!t]
\caption{Network sensitivity analysis across different structures and fault types.}
\label{tab:network_sensitivity} 
\centering
\begin{threeparttable}
  \begin{tabular}{@{}l l c c c c c c c c@{}}
    \toprule
\textbf{Faulted Line} & \textbf{Structure} & \textbf{Fault Type} & \bm{$R_f${(\si{\ohm})}} & \bm{$D_{\!M}^2$} & \textbf{Detection Time (ms)} & [\bm{$z_a$}, \bm{$z_b$}, \bm{$z_c$}, $g_0^*$] & \textbf{Predicted Fault Type}\\
    \midrule
    Line 07-08 & grid-connected  & \textit{ag}            & 50     & $\approx 10^4$  & 20  & [155, 0, 0.5, 100]   & \textit{ag}  \\ 
    Line 03-04 & islanded        & \textit{bcg}           & 100    & 150             & 18  & [2.9, 102, 5, 100] & \textit{bcg} \\
    Line 01-02 & grid-connected  & \textit{bg}            & 100    & $\approx 10^4$  & 10  & [0, 55.3, 4.1, 130]      & \textit{bg}  \\
    Line 09-10 & islanded        & \textit{ab}            & -      & $\approx 10^4$  & 2   & [120, 46, 0.1, 10]     & \textit{ab}  \\
    Line 12-13 & grid-connected  & \textit{abc}           & -      & $\approx 10^4$  & 12  & [155, 110, 80, 11]   & \textit{abc} \\
    Line 13-14 & grid-connected  & \textit{ac}            & -      & $\approx 10^4$  & 2   & [65, -0.1, 65, 12]    & \textit{ac} \\ 
    Line 05-06 & islanded        & \textit{bcg}           & 150    & $\approx 10^4$  & 14  & [2.3, 62.7, 58, 80]      & \textit{bcg} \\
    Line 02-03 & grid-connected  & \textit{cg}            & 50     & $\approx 10^4$  & 6   & [-0.3, 0, 120, 125]  & \textit{cg} \\
    \bottomrule
  \end{tabular}
\end{threeparttable}
\end{table*}

\subsection{Topology Change}
\label{subsec:topology-change}

\subsubsection{Grid-connected Mode}
\label{subsubsec:grid-connected-mode}
\begin{figure}[htbp]
  \centering
  \includegraphics[width=\linewidth]{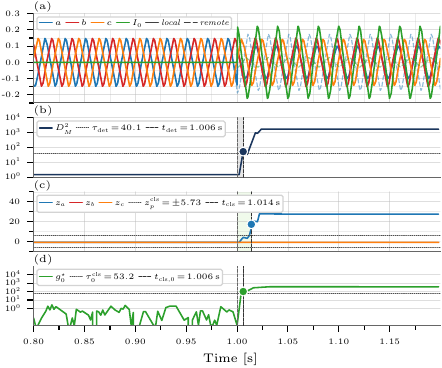}
  \caption{Time-domain response of detection and classification for a 50~\(\Omega\) LG fault on line 01–02 (grid-connected). The fault begins at \(t=1.0\)~s. Detection occurs at \(t=1.006\)~s as \(D_{\!M}^2\) exceeds \(\tau_{\mathrm{det}}\), then it is classified as \textit{ag} by the \textit{relative jump test} in the phase-a G-statistic and by the zero-sequence statistic \(g_0^{*}(t)\) crossing its absolute threshold.}

  \label{fig:grid_diagram}
\end{figure}

Under grid-connected conditions, a single-line-to-ground fault (LG), with a fault resistance of \SI{50}{\ohm} was applied to line 01-02 at $t = \SI{1.0}{s}$. Fig.~\ref{fig:grid_diagram}  illustrates the temporal evolution of the squared Mahalanobis distance, $D_{\!M}^2$. Approximately  \SI{6}{\milli\second} after fault inception, $D_{\!M}^2$ exceeds $\tau_{\text{det}}$, triggering the protection scheme.
Immediately upon detection, the classification logic is invoked. The system identifies the event as an \textit{ag} fault based on the two test: the rapid increase in the G-statistic for phase a (\textit{Relative jump test}), and the zero-sequence statistic, $g^*_{0} (t)$, exceeding its $\tau_0^{\mathrm{cls}}$. This dual-criterion approach ensures a swift classification even before the phase's z-score stabilizes above its own threshold.

\subsubsection{Islanded Mode}
\label{subsubsec:islanded_diagram}
\begin{figure}[htbp]
  \centering
  \includegraphics[width=\linewidth]{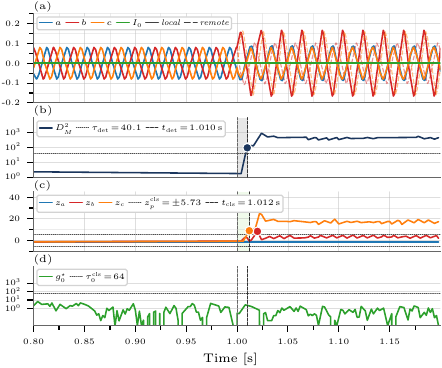}
\caption{Time-domain response of detection and classification for an LL fault on line 03–04 (islanded). The fault occurs at \(t=1.0\)~s. \(D_{\!M}^{2}\) crosses \(\tau_{\mathrm{det}}\) at \(t=1.010\)~s. Then \(z_b(t)\) and \(z_c(t)\) cross their respective thresholds, correctly identifying the event as \textit{bc}.}

\label{fig:islanded_diagram}
  \label{islanded_diagram}
\end{figure}

Under islanded conditions, a line-to-line (LL) fault was applied to line 03-04 at $t = \SI{1.0}{s}$. Fig.~\ref{fig:islanded_diagram} shows the temporal evolution of the squared Mahalanobis distance, $D_{\!M}^2$. Just \SI{10}{\milli\second} after fault inception, the $D_{\!M}^2$ statistic sharply rises past $\tau_{\text{det}}$, triggering the protection scheme. Immediately following this detection, the classification module is activated. The ${z_b(t)}$ and ${z_c(t)}$, are flagged by the classifier (through the \textit{absolute test}), correctly identifying the event as a \textit{bc} fault.

Additional results are presented in Table~\ref{tab:network_sensitivity}, which details the scheme's performance across various scenarios. For each case, it reports the key detection and classification metrics, including the peak $D_{\!M}^2$, ${z_p}$ , and the $g_0^*$, along with the resulting detection time. These results validate the scheme’s robust performance, demonstrating both rapid, sub-cycle detection and accurate fault-type discrimination in islanded and grid-connected modes.

\subsection{Combined Operational Disturbances and Faulted Case}
\label{subsec:combined-disturbances}
\begin{figure}[t!]
  \centering
  \includegraphics[width=\linewidth]{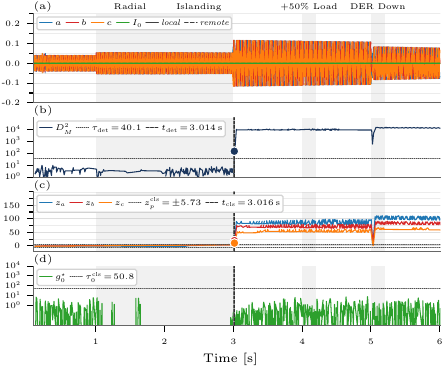}
  \vspace{-0.1in}
  \caption{Response of \(D_{\!M}^{2}\) to a sequence of challenging events on feeder 03–08. After reconfiguration and islanding, an LLL fault is initiated at \(t=3.0\)~s. The index crosses \(\tau_{\mathrm{det}}\) and remains stable, even as a \(50\%\) load increase (at \(t=4.0\)~s) and DER output reduction (at \(t=5.0\)~s) create more adverse operating conditions.}

  \label{fig:integrated_fault0308}
\end{figure}

To evaluate the robustness of the detector under combined dynamic and faulted conditions, a sequential test was conducted on line 03-08, which supplies load L8 and integrates DER2 (PV). The system operated in ring configuration; at \(t = 1.0~\mathrm{s}\) it was switched to radial mode by opening S2, and at \(t = 2.0~\mathrm{s}\) the microgrid was islanded from the grid by opening CB1. At \(t = 3.0~\mathrm{s}\), an internal LLL fault was applied on line 03-08. While the fault persisted, further operational changes were introduced: the load at bus~8 (L8) was increased by 50\% at \(t = 4.0~\mathrm{s}\), and DER2 (PV) output was reduced by 50\% at \(t = 5.0~\mathrm{s}\), emulating a loss of local generation support. Despite these compound disturbances and reduced fault-current levels during islanded operation, as illustrated in Fig.~\ref{fig:integrated_fault0308}, \(D_{\!M}^2\) promptly exceeded \(\tau_{\mathrm{det}}\) at fault inception and remained above it, demonstrating the detector’s reliability under complex, time-varying conditions.

\subsection{Communication Delay}
\begin{figure}[htbp]
  \centering
  \includegraphics[width=\linewidth]{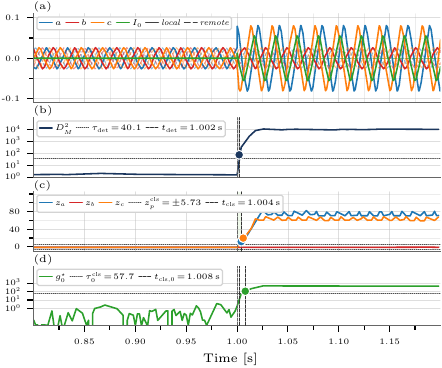}

 \caption{Time-domain response of fault detection and classification for an LLG fault on feeder 04–05 with a \SI{10}{ms} communication delay. The fault occurs at \(t=1.0\)~s. \(D_{\!M}^{2}\) crosses \(\tau_{\mathrm{det}}\) at \(t=1.002\)~s. Then \(z_b(t)\) and \(z_c(t)\) cross their respective thresholds, correctly identifying the event as a \textit{bcg} fault.}

  \label{fig:comm_delay}
\end{figure}

To evaluate the effect of communication latency on the protection scheme, a fixed one-way delay of \SI{10}{\milli\second} is introduced on the remote-end current measurements of line 04-05 operating in islanded mode. 
An LLG fault is applied to the line at $t=\SI{1.0}{s}$ and maintained for the rest of the simulation. 
As shown in Fig.~\ref{fig:comm_delay}, the time-domain response of $D_{\!M}^2$ indicates that, despite the \SI{10}{ms} communication delay, the detection index exceeds the threshold $\tau_{\det}$ in only \SI{2}{\milli\second}. Additionally, $z_b(t)$ and $z_c(t)$, as well as $g_0^*(t)$, correctly identifying the event as a \textit{bcg} fault.

\subsection{High-Impedance Fault Under Noisy Condition}
\begin{figure}[htbp]
  \centering
  \includegraphics[width=\linewidth]{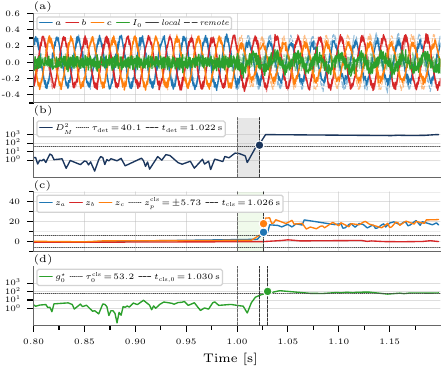}
  \caption{Time-domain response of \(D_{\!M}^{2}\) for a \SI{250}{\ohm} LLG fault on line 01–02, in the presence of additive white Gaussian noise with an SNR of \SI{20}{dB}. The fault occurs at \(t=1.0\)~s. \(D_{\!M}^{2}\) crosses \(\tau_{\mathrm{det}}\) at \(t=1.022\)~s. Then \(z_a(t)\), \(z_c(t)\), and \(g_0^{*}(t)\) cross their respective thresholds, correctly identifying the event as an \textit{acg} fault.}
  \label{fig:High-Impedance Fault Under Noisy Condition}
\end{figure}

High-Impedance Fauls are characterized by fault currents that barely exceed normal load levels, often accompanied by intermittent arcing, which makes them inherently difficult to detect. Measurement noise further masks these low-amplitude signatures, increasing the risk of missed detections\cite{ref4}. To evaluate the scheme's resilience, a high-resistance LLG fault of \SI{250}{\ohm} is applied to line 01-02 at \(t=1.0\)~s, while additive white Gaussian noise with a SNR of \SI{20}{dB} is superimposed on both local and remote current measurements. Fig.~\ref{fig:High-Impedance Fault Under Noisy Condition} shows the time-domain evolution of $D_{\!M}^2$. Despite the low fault-current amplitude and severe noise, the index still exceeds its $\tau_{\mathrm{det}}$ within \SI{22}{\milli\second}, demonstrating robust detection without false trips.

\subsection{Overall Performance Summary}
The efficacy of the proposed protection scheme is evaluated using standard performance metrics that capture both fault detection and classification. All metrics are aligned with the actionable timing of a practical relay, with decisions timestamped at the end of the corresponding analysis window.

For fault detection, three complementary aspects are considered: sensitivity, security, and speed. Sensitivity is quantified by the detection probability ($P_D$), defined as the fraction of fault events correctly identified. Security is measured by the false alarm rate (FAR), i.e., the proportion of healthy windows incorrectly flagged. Detection speed is captured by the detection delay ($T_D$), defined as:
\begin{equation}
    T_D = \min \left\{ t_w : D_{\!M}^2(t_w) > \tau_{\text{det}} \right\} - t_f.
\end{equation}
where $t_w$ represents the actionable timestamp of each analysis window (typically the window's end time), and $t_f$ is the known ground-truth time of the fault's inception.

\begin{table*}[htbp]
\centering
\caption{Overall performance summary for fault detection and classification under varying noise conditions}
\label{tab:overall_performance_summary}
\sisetup{
  round-mode=places, round-precision=2,
  detect-weight=true, detect-inline-weight=math
}

\begin{tabular*}{\textwidth}{@{\extracolsep{\fill}} l
  S[table-format=2.2]  
  S[table-format=1.1]  
  S[table-format=2.2]  
  S[table-format=2.2]  
  S[table-format=2.2]  
@{}}
\toprule
& \multicolumn{3}{c}{\textbf{Fault Detection}} & \multicolumn{2}{c}{\textbf{Fault Classification}} \\
\cmidrule(lr){2-4} \cmidrule(lr){5-6}
\textbf{Scenario} & {Avg. Time (\si{ms})} & {FAR (\%)} &
{$P_D$ (\%)} & {Accuracy (\%)} & {Macro F1 (\%)} \\
\midrule
Clean (no noise) & 12.57 & 1.2 & 97.78 & 97.43 & 97.88 \\
40 dB SNR        & 12.92 & 2.1 & 98.11 & 96.33 & 97.33 \\
30 dB SNR        & 13.90 & 2.8 & 98.33 & 95.95 & 95.55 \\
20 dB SNR        & 18.84 & 6.7 & 99.55 & 93.83 & 93.33 \\
\bottomrule
\end{tabular*}

\medskip
\end{table*}

\begin{table*}[!t]
  \centering
   \caption{Comparative Overview of Protection Methods}
  \label{tab:comparison}
  \scriptsize
   \begin{threeparttable}
     \begin{tabularx}{\textwidth}{@{}l l *{10}{>{\centering\arraybackslash}X}@{}}
      \toprule
       \multirow{1}{*}{\textbf{Metric}} 
         & \multirow{1}{*}{\textbf{}} 
         & \multicolumn{1}{c}{\textbf{Proposed}} 
         & \multicolumn{1}{c}{\cite{ref17}} & \multicolumn{1}{c}{\cite{ref24}} & \multicolumn{1}{c}{\cite{ref22}} & \multicolumn{1}{c}{\cite{ref21}} 
         & \multicolumn{1}{c}{\cite{ref19}} & \multicolumn{1}{c}{\cite{ref20}} & \multicolumn{1}{c}{\cite{ref23}} & \multicolumn{1}{c}{\cite{ref18}} & \multicolumn{1}{c}{\cite{ref28}} \\
       \midrule
      \multirow{5}{*}{\textbf{Functional Performance}}
         & Grid-Tied Fault Detection & \cmark & \xmark & \cmark & \cmark & \cmark & \cmark & \cmark & \cmark & \cmark & \cmark \\
         & Islanded Mode Fault Detection & \cmark & \cmark & \cmark & \cmark & \cmark & \cmark & \cmark & \cmark & \cmark & \cmark \\
         & Faulted Phase Identification & \cmark & \cmark & \cmark & \cmark & \cmark & \cmark & – & \cmark & \cmark & \cmark \\
         & HIF Consideration & \cmark & \cmark & \cmark & \cmark & \cmark & \cmark & \cmark & \cmark & \cmark & \cmark \\
       \midrule
      \multirow{3}{*}{\textbf{Technical Limits}}
         & Max $R_f$ Specified (\si{\ohm}) & 250 & 200 & 100 & \xmark & \xmark & \xmark & 50 & 170 & 100 & \xmark \\
         & Noise Resilience (Min SNR, dB) & 20 & 40 & 25 & 30 & 25 & \xmark & \xmark & 25 & 30 & – \\
         & Comm. Delay Immunity (ms) & 10 & \cmark & \xmark & \xmark & \xmark & \xmark & \xmark & \xmark & 8.33 & \xmark \\
       \midrule
       \multirow{2}{*}{\textbf{Deployment Practicality}}
         & Threshold Configuration Method & \textbf{A} & E & E & E & E & E & E & E & E & Z \\
         & Adaptability to Load/DG Variations & \cmark & \cmark & \cmark & \cmark & \cmark & \cmark & \xmark & \xmark & \cmark & \cmark \\
       \bottomrule
     \end{tabularx}
     \begin{tablenotes}[flushleft]\footnotesize
       \item \textbf{Key:}  Threshold Method: {A = Analytical/Statistical}, E = Empirical/Heuristic, Z = Zero-Crossing Based.
     \end{tablenotes}
   \end{threeparttable}
\end{table*}

\subsubsection{Fault Detection Performance}
\begin{figure}[htbp]
    \centering
    \includegraphics[width=0.85\linewidth]{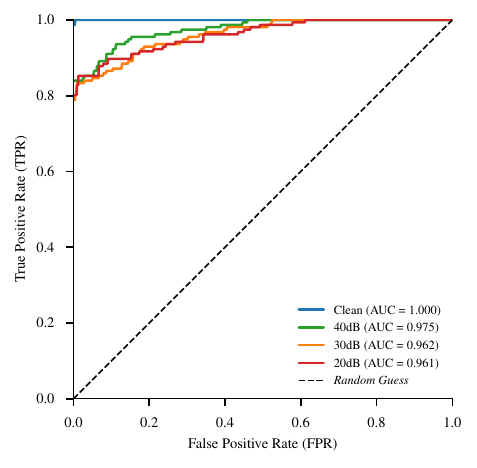}
    \caption{ROC curves for the fault detector on a representative feeder (line 10–11) under varying noise conditions. The high Area Under the Curve (AUC) values demonstrate robust detection performance even at high SNR.}

    \label{fig:roc_example}
\end{figure}

Table~\ref{tab:overall_performance_summary} summarizes detection performance across all feeders. Even under severe noise (20 dB SNR), the detector maintains a high $P_D$ of 99.55\% with an average delay of \SI{18.84}{\milli\second}, while FAR remains negligible. These results highlight the robustness of the Mahalanobis-based scheme and the reliability of its $\chi^2$ thresholding. To offer a more granular view, Fig.~\ref{fig:roc_example} presents the Receiver Operating Characteristic (ROC) curves for a representative line, line 10-11. The plot visually confirms the trend observed in the aggregate data, showing a graceful degradation in performance as noise increases. The Area Under the Curve (AUC), a holistic measure of detectability, remains high even at 20 dB SNR, indicating the detector's strong discriminative power between faulty and healthy states throughout the tested conditions.

\subsubsection{Fault Classification Performance}
\label{ssec:Fault_Classification_Performance}
Following detection, the classifier identifies the fault type. Table~\ref{tab:overall_performance_summary} reports precision, recall, and F1-scores aggregated across all feeders. Results show high accuracy that degrades gradually with noise, with macro-averaged F1-scores confirming balanced performance across classes.

\begin{figure}[htbp]
    \centering
    \includegraphics[width=0.97\linewidth]{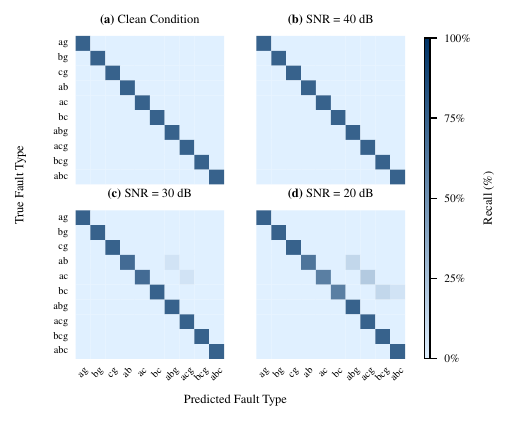}
    \caption{Confusion matrices for the fault classifier on a representative feeder (line 12–13) under (a) clean conditions, (b) \SI{40}{dB} SNR, (c) \SI{30}{dB} SNR, and (d) \SI{20}{dB} SNR. The classifier remains highly accurate in the clean case and shows predictable confusion patterns under noise.}
    \label{fig:cm_example}
\end{figure}

To illustrate the nature of classification errors, Fig.~\ref{fig:cm_example} displays the confusion matrices for another representative feeder, line 12-13, under both clean and severe noise conditions. Under clean conditions (Fig.~\ref{fig:cm_example}(a)), the classification is nearly flawless, with entries heavily concentrated along the main diagonal. In the presence of 20 dB noise (Fig.~\ref{fig:cm_example}(d)), a slight increase in misclassifications is observed. Notably, the confusion primarily occurs between fault types with similar signatures, such as LLG (e.g., abg) faults. This behavior is expected and highlights the inherent challenges of precise classification in low SNR environments.

\section{Comparative Analysis}
To contextualize the contributions of this work, we compare our proposed scheme with several recent signal processing and statistical methods for microgrid protection. Table~\ref{tab:comparison} provides a detailed overview based on characteristics.

Many contemporary protection schemes, particularly those based on time-frequency transforms like wavelet \cite{ref21}, \cite{ref22}, S-transform \cite{ref23}, or empirical mode decomposition \cite{ref19}, \cite{ref20}, rely heavily on empirically tuned or heuristic thresholds. While effective under specific conditions, this dependency can limit their robustness and complicates field deployment, often requiring extensive re-tuning for different network topologies or operational states.

In stark contrast, our methodology introduces a statistically principled framework. The $\tau_{\mathrm{det}}$ is not an empirical value, but is analytically derived from the theoretical $\chi^2$ distribution of the Mahalanobis distance. This provides direct control over the false alarm rate and ensures consistent, predictable performance across diverse scenarios, a significant advantage in terms of reliability and ease of implementation.

Furthermore, our scheme demonstrates superior performance in several key areas. It reliably detects HIFs up to \SI{250}{\ohm}, a capability that exceeds many existing methods. While our HIF model uses a standard resistive representation, this approach effectively validates the detector's sensitivity to the low-magnitude fault currents characteristic of such events. The method also maintains high robustness against severe measurement noise, up to \SI{20}{dB} SNR, and tolerates significant communication delays, up to \SI{10}{\milli\second}, critical factors for practical deployment that are often overlooked in the literature.

Collectively, these attributes position our approach as a robust, transparent, and computationally efficient solution that bridges the gap between theoretical rigor and practical applicability for next-generation microgrid protection.

\section{Conclusion}
\label{sec:conclusion}
This paper has introduced a statistically adaptive differential protection framework for AC microgrids. The core of the methodology is a multivariate fault detector that leverages the $D_{\!M}^2$ to measure deviations from a well-characterized healthy state. This approach, built upon KL divergence implemented via a Bartlett-corrected G-statistic, is distinguished by its principled thresholding, which is derived directly from the theoretical $\chi^2$ distribution to provide precise control over the false alarm rate.

Extensive simulations on a modified CIGRE 14-bus microgrid confirmed the scheme's high efficacy. It achieved an overall detection probability (\(P_D\)) exceeding \SI{99}{\percent} and a classification F1-score of \SI{97}{\percent} across a wide range of challenging conditions, including high-impedance faults up to \SI{250}{\ohm} and severe noise down to \SI{20}{dB} SNR. The method consistently delivered sub-cycle average detection delays with a minimal computational footprint, well within the capacity of modern protective relays.

These findings highlight the potential of this framework as a transparent, reproducible, and computationally efficient solution for the next generation of microgrid protection systems. Future work will focus on three key directions: 1) Hardware Implementation and Validation through deployment on FPGA-based platforms and comprehensive Hardware-in-the-Loop (HIL) testing incorporating realistic current transformer saturation and IBR control dynamics; 2) Enhanced Fault Classification by exploring lightweight, interpretable machine learning models to improve discrimination in ambiguous, high-resistance scenarios; and 3) Extension to Complex Topologies, adapting the core statistical framework to protect meshed or multi-voltage-level active distribution networks.

\bibliographystyle{IEEEtran}

\end{document}